\documentclass[a4paper,11pt]{article}

\usepackage{jheppub}
\usepackage[T1]{fontenc}
\usepackage{amsmath,amssymb,mathtools}
\usepackage{booktabs}
\usepackage{array}
\usepackage{microtype}

\newcommand{\Lattice}{\Gamma}

\newcommand{\Cone}{\operatorname{Cone}}

\newcommand{\R}{\mathbb{R}}
\newcommand{\Z}{\mathbb{Z}}
\newcommand{\cT}{\mathcal{T}}

\title{Intersection Bounds for BPS Strings in Six-Dimensional Supergravity}

\author[a,b]{Hee-Cheol Kim}
\emailAdd{heecheol@postech.ac.kr}
\author[c]{Kai Xu}
\emailAdd{kaixu1996@gmail.com}

\affiliation[a]{Department of Physics, POSTECH, Pohang 37673, Korea}
\affiliation[b]{Asia Pacific Center for Theoretical Physics, Postech, Pohang 37673, Korea}
\affiliation[c]{Jefferson Physical Laboratory, Harvard University, Cambridge, MA 02138, USA}

\abstract{In six-dimensional $\mathcal{N}=(1,0)$ supergravity, the structure of tensor moduli space is governed by primitive BPS string charges known as BPS generators and their intersection pairing. We derive bounds on the intersection numbers of these generators from a purely effective field theory (EFT) perspective. Although gauge anomaly cancellation constrains intersections between generators supporting gauge algebras, bounds for E-strings intersecting generators with self-intersection numbers $-2$ and $-3$ have previously remained incomplete. We show that the Zariski decomposition, interpreted as the charge lattice counterpart of the attractor mechanism, together with current algebra embeddings on the E-string worldsheet theory, yields strong universal bounds on these intersection numbers. These results establish the finiteness of tensor charge intersection numbers up to duality. The underlying structure was identified through AI-guided investigation and is proven here analytically using EFT arguments.}

\begin{document}
\hypersetup{pageanchor=false}
\maketitle

\hypersetup{pageanchor=true}
\flushbottom

\section{Introduction}
\label{sec:introduction}

Understanding which low-energy effective field theories (EFTs) admit a consistent ultraviolet completion in quantum gravity is one of the central goals of the Swampland program~\cite{Vafa:2005ui,Agmon:2022thq}. Six-dimensional $\mathcal{N}=(1,0)$ supergravity has long served as one of the most fertile testing grounds for this program. Its low-energy spectrum and interactions are tightly constrained by anomaly cancellation~\cite{Green:1984bx,Sagnotti:1992qw}, while a broad class of consistent models can be constructed explicitly in string theory, most notably through F-theory compactifications~\cite{Vafa:1996xn,Morrison:1996na,Morrison:1996pp}. As a result, six-dimensional supergravity has become an exceptional laboratory in which one can compare purely low-energy consistency conditions with explicit constructions from quantum gravity. This interplay has led to a series of remarkable results over the past decade, and has demonstrated that many powerful constraints on the space of consistent theories can be established using EFT considerations.

A striking recent development is the discovery of a new universal structure of six-dimensional supergravity from effective field theory arguments alone \cite{Kim:2019vuc,Kim:2024eoa}. Assuming only the completeness of the classification of 6d superconformal field theories (SCFTs) and little string theories (LSTs)~\cite{Heckman:2015bfa,Bhardwaj:2015xxa,Bhardwaj:2015oru,Bhardwaj:2018jgp,Bhardwaj:2019hhd}, together with the existence of tensionless BPS strings at every boundary of tensor moduli space, it was shown that every consistent $6d~\mathcal{N}=(1,0)$ supergravity theory contains only finitely many massless fields~\cite{Kim:2024eoa}. In particular, the number of tensor multiplets was proven to satisfy the universal bound $T\le 193$, which provides the first purely EFT proof of the finiteness of the massless spectrum.  Furthermore, every consistent tensor charge lattice was shown to be realizable as the topology of a Kähler surface obtained from $\mathbb{P}^2$, a Hirzebruch surface $\mathbb{F}_n$, or one of their blowups~\cite{Kim:2024eoa}. These results strongly suggest that the tensor sector of 6d supergravity possesses a much richer universal structure than had previously been recognized.

Despite this progress, one important aspect of the tensor sector has remained largely unexplored from the EFT perspective. The tensor charge lattice is characterized not only by types of primitive string charges it contains, but also by their mutual intersection pairing. While the topology of the lattice itself is now understood from the topology of a K\"ahler surface as proven in~\cite{Kim:2024eoa}, no general EFT derivation of universal bounds on the mutual intersection numbers between BPS string charges has been available. This issue is not merely technical. Universal intersection bounds play an essential role in the classification of 6d supergravity theories, since they determine the structure of possible tensor moduli space that must be scanned in any systematic manner. In particular, the absence of such bounds has been one of the main obstacles in ongoing classification programs, for example in \cite{Hamada:2026zta}, including the complete enumeration of tensor bases. Together with the independent problem of unbounded Abelian gauge charges~\cite{Taylor:2018khc}, the lack of EFT bounds on tensor intersection numbers has prevented a purely low-energy proof of the finiteness of the massless spectrum.

The problem of bounding tensor intersection numbers can be formulated more precisely. When two tensor charges both support gauge algebras, their mutual intersection numbers are completely fixed by gauge anomaly cancellation and hence follow directly from low-energy consistency~\cite{Green:1984bx,Sagnotti:1992qw}. However, this argument ceases to apply once one of the two charges does not carry a gauge algebra.  The canonical example is the E-string charge, which has self-intersection $-1$ and is called $(-1)$-charge. Although its worldsheet theory possesses an $(E_8)_1$ current algebra~\cite{Minahan:1998vr,Kim:2014dza}, it does not support a 6d gauge algebra. In particular, intersections between the E-string and tensor charges supporting small gauge algebras, such as $\mathfrak{su}(2)$ or $\mathfrak{su}(3)$, have remained essentially unconstrained from the EFT perspective. By contrast, F-theory compactifications predict strong geometric restrictions on these intersection numbers, and explicit bounds were obtained in~\cite{Morrison:2012np}. This discrepancy indicates that the current EFT understanding of the tensor sector is still incomplete and suggests that additional universal structures remain to be uncovered.

The main goal of this paper is to derive universal bounds on these previously uncontrolled tensor intersections using only minimal EFT consistency conditions. Our starting point is the observation that the gravitational string charge, denoted by $b_0$, admits a natural decomposition into positive and negative components, closely analogous to the Zariski decomposition of divisors on algebraic surfaces~\cite{Zariski:1962}. While Zariski decomposition is traditionally formulated in algebraic geometry, we show that an equivalent structure arises intrinsically within the tensor charge lattice itself. From the physical viewpoint, this decomposition is naturally interpreted as the lattice realization of the black string attractor mechanism~\cite{Ferrara:1995ih,Ferrara:1996dd,Denef:2000nbc,HetLam:2018yba}. The negative part identifies the tensor charges that drive the attractor flow, whereas the positive part has non-negative pairing with every populated BPS string charge. Once this structure is recognized, the unit intersection relation between the gravitational string charge and a primitive $(-1)$-charge immediately translates into upper bounds on previously uncontrolled intersection numbers.

The resulting bounds are remarkably strong. For a primitive tensor charge with self-intersection $-3$, which we refer to as a $(-3)$-charge, we prove that its intersection with a primitive $(-1)$-charge is universally bounded by three.  The same method extends to $(-2)$-charges appearing in the tensor chains $(-2,-3)$, $(-2,-2,-3)$, and $(-2,-3,-2)$ introduced in~\cite{Morrison:2012np,Heckman:2015bfa}, providing the universal intersection bound $7$. The only genuinely exceptional case is an isolated $(-2)$-charge supporting a gauge algebra $G$. Here the attractor decomposition becomes degenerate, and consequently no finite bound follows from the attractor mechanism alone. We therefore employ another physical argument. The relevant constraint we find in this case instead comes from the worldsheet theory on an E-string. The EFT consistency requires, when an isolated $(-2)$-charge with a gauge algebra $G$ intersects the E-string, this algebra to embed into the level-one $E_8$ current algebra in the E-string worldsheet CFT~\cite{Kim:2019vuc,Kim:2016foj,Shimizu:2016lbw}. This immediately produces the finite universal bound $1240$~\cite{Dynkin:1957,Panyushev:2009,Panyushev:2014} on the intersection number, although this bound is considerably weaker than those obtained in the other cases. Taken together, these results eventually establish the finiteness of the tensor charge intersection numbers up to duality acting on the charge lattice.

Finally, we briefly comment on the origin of the ideas developed in this work. The central organizing principle, the lattice Zariski decomposition and its application to tensor intersection bounds, was not discovered through a direct analytic derivation in field theory. Rather, it emerged during an AI-assisted exploration of the tensor charge lattice, initiated by the question of whether universal intersection bounds could be derived solely from the effectiveness of the gravitational string charge and the mutual non-negativity of primitive BPS string charges. Repeated interactions with an AI agent (Claude Fable 5) suggested the underlying structure, which was subsequently reformulated into rigorous mathematical statements. Every result presented in this work is established independently by analytic EFT arguments.

The paper is organized as follows. In Section~\ref{sec:setup}, we review the 6d supergravity, tensor charge lattice, primitive BPS charges, and the EFT assumptions used throughout the paper. In Section~\ref{sec:zariski}, we construct the lattice Zariski decomposition of the gravitational string charge and explain its interpretation as the lattice analogue of the black string attractor mechanism. Section~\ref{sec:masterbound} derives the general intersection bound for $(-1)$-charges via the Zariski decomposition and the current algebra embedding. We conclude in Section~\ref{sec:discussion} with a discussion of the AI-guided discovery process and the implications for the classification of 6d supergravity theories and possible future directions.

\section{Six-dimensional $\mathcal N=(1,0)$ supergravity}
\label{sec:setup}

In this section, we briefly review the aspects of 6d $\mathcal N=(1,0)$ supergravity that will be needed throughout this paper. We also summarize the key results of Ref.~\cite{Kim:2024eoa} that will serve as essential ingredients in our proofs.

\subsection{Massless spectrum, tensor charge lattice, and anomaly cancellation}
\label{subsec:eft-review}

The massless spectrum of a six-dimensional $\mathcal N=(1,0)$ supergravity theory consists of a gravity multiplet together with tensor, vector, and hypermultiplets. The gravity multiplet contains the graviton $g_{\mu\nu}$ and a self-dual two-form field $B^+_{\mu\nu}$, while each tensor multiplet contains an anti-self-dual two-form field $B^-_{\mu\nu}$ and a real scalar. The $T$ real scalar fields in the tensor multiplets parametrize the tensor moduli space, which is locally given by the coset $SO(1,T)/SO(T)$~\cite{Nishino:1984gk,Nishino:1986dc,Romans:1986er}.

Because the two-form fields are (anti-)self-dual, their electric and magnetic string charges take values in the same charge lattice. Dirac quantization and consistency of the self-dual fields require this lattice to be an integral unimodular lattice \cite{Seiberg:2011dr},
\begin{equation}
 \Lattice\subset\R^{1,T},
 \qquad
 v\cdot w=\Omega_{\alpha\beta}v^\alpha w^\beta\in\Z,
 \qquad
 \operatorname{sign}(\Omega)=(1,T),
 \label{eq:charge-lattice}
\end{equation}
where $\Gamma=\Gamma^*$ and $\Omega_{\alpha\beta}$ is the tensor intersection form on the tensor charge lattice.

The chiral massless fields generate gravitational, gauge, and mixed anomalies.  These anomalies are canceled by the Green-Schwarz-Sagnotti mechanism when the one-loop anomaly polynomial $I_8$ factorizes as \cite{Green:1984sg,Green:1984bx,Sagnotti:1992qw,Erler:1993zy}
\begin{equation}
 I_8=\frac12\Omega_{\alpha\beta}X_4^\alpha X_4^\beta,
 \qquad
 X_4^\alpha
 =-\frac12 b_0^\alpha\operatorname{tr}R^2
 +\frac14\sum_i b_i^\alpha\frac{2}{\lambda_i}
 \operatorname{tr}F_i^2 .
 \label{eq:anomaly-factorization}
\end{equation}
The vector $b_0$ is the gravitational anomaly vector, $b_i$ is the anomaly vector for the gauge factor $\mathfrak{g}_i$, and $\lambda_i$ is its standard group-normalization constant. These vectors are elements of the tensor charge lattice after the appropriate quantization conditions are imposed \cite{Seiberg:2011dr,Monnier:2017oqd}. 

The factorization condition above imposes stringent constraints on the massless spectrum. Among these constraints, the mixed gauge anomaly cancellation condition is particularly important for the present work.  If two tensor charges support gauge algebras $\mathfrak g_i$ and $\mathfrak g_j$, their mutual intersection is completely determined by the spectrum of bifundamental matter charged under both gauge groups as
\begin{equation}
 b_i\cdot b_j
 =2\lambda_i\lambda_j\sum_{R,S}n_{RS}^{ij}A_R^iA_S^j,
 \qquad i\neq j,
 \label{eq:mixed-gauge-anomaly}
\end{equation}
with $\operatorname{tr}_{R}F_i^2=A_R^i\operatorname{tr}F_i^2$ where $n_{RS}^{ij}$ is the number of hypermultiplets transforming in $(R,S)$-representation under $\mathfrak{g}_i\times \mathfrak{g}_j$. Thus, whenever both tensor charges support gauge algebras, their intersection number is fixed by anomaly cancellation. This relation, however, does not provide an analogous constraint when one of the tensor charges supports no gauge algebra, as in the case of the E-string or the M-string. It is precisely this gap that gives rise to the missing intersection bounds addressed in this paper.

The tensor branch of a 6d $(1,0)$ supergravity theory contains half-BPS strings carrying tensor charges $Q\in\Gamma$~\cite{Seiberg:2011dr,HetLam:2018yba,Kim:2019vuc}, whose tensions are determined by
\begin{equation}
\mathfrak{T}_Q\sim J\cdot Q,
\end{equation}
where $J$ denotes the vacuum expectation values of the $T$ tensor multiplet scalars. Of particular interest are the strings that become tensionless at the boundary of tensor moduli space, namely those satisfying $J\cdot Q=0$. The mutual intersection pairing of these BPS strings governs the structure of the tensor moduli space and will be the central object of this work. Recent developments in the classification of 6d SCFTs, little string theories, and the boundedness of 6d supergravity have uncovered several universal properties of these BPS generators. Since these results provide the physical foundation for our analysis, we briefly review them below.

\subsection{BPS cone and BPS generators}
\label{subsec:bps-cone}

A major recent advance was achieved in~\cite{Kim:2024eoa}, where a new universal structure of six-dimensional $\mathcal N=(1,0)$ supergravity was derived from purely effective field theory arguments. The analysis is based on two assumptions. First, the existing classification of 6d SCFTs and little string theories, developed for example in ~\cite{Morrison:2012np,Heckman:2015bfa,Bhardwaj:2015xxa,Bhardwaj:2015oru,Bhardwaj:2018jgp,Bhardwaj:2019hhd}, is complete. Second, every boundary of the tensor moduli space hosts tensionless BPS strings. Together, these assumptions imply that every tensionless string sector appearing at a boundary of tensor moduli space must belong to the known classification.

The physical BPS strings generate a convex cone inside $\Lattice\otimes\mathbb R$, which we call the BPS cone~\cite{Kim:2024eoa}. We denote its primitive charges by $C_i$ and refer to them, or equivalently to the associated strings, as primitive BPS generators. Positivity of their tensions defines the dual tensor cone,
\begin{equation}
 \cT=\left\{
 J\in\R^{1,T}\ \middle|\
 J^0>0,\;
 J^2\ge0,\;
 J\cdot C_i\ge0
 \text{ for all generators }C_i
 \right\}\ .
\end{equation}
The tensor moduli space is the unit slice of this cone satisfying $J^2=1$. Consequently, the primitive BPS generators determine the walls, faces, and asymptotic directions of the tensor cone. Its boundaries correspond to loci where one or more generators satisfy $J\cdot C_i=0$ and become tensionless. Since only non-spacelike charges can be orthogonal to the timelike vector $J$ in a Lorentzian space of signature $(1,T)$, every primitive BPS generator satisfies
\begin{equation}
C_i^2\le0 \ .
\end{equation}


Since every primitive BPS generator becomes tensionless at some boundary of the tensor moduli space, all such generators have been classified through the classification of 6d SCFTs and little string theories~\cite{Heckman:2015bfa,Bhardwaj:2015xxa,Bhardwaj:2015oru,Bhardwaj:2018jgp,Bhardwaj:2019hhd}. Finite-distance boundaries are described by local SCFT strings, whereas infinite-distance boundaries are associated with little strings or critical strings. Besides instantonic strings with charges $b_i$ supporting non-Abelian gauge algebras, the classified primitive generators include BPS generators that carry no gauge algebra:
\begin{align}
     &1)\ e^2=-1,\quad b_0\cdot e=1\, ,\qquad 2)\ C^2_{-2}=-2,\quad b_0\cdot C_{-2}=0\, ,\nonumber\\
     &3)\ C^2_{\rm het}=0,\quad b_0\cdot C_{\rm het}=2\, ,\qquad 4)\ C^2_{\rm II}=0,\quad b_0\cdot C_{\rm II}=0,
\end{align}
corresponding to the E-string, M-string, and the critical heterotic and Type II strings, respectively~\cite{Minahan:1998vr,Kim:2014dza,Haghighat:2013gba,Kim:2024eoa}. A summary of all primitive BPS generators is given in Table~\ref{tb:generators}.

\begin{table}[t]
\centering
\begin{tabular}{|c|l|c|c|c|}
    \hline
    $\mathfrak{g}_i$ & $H_i$ & $b_i^2$ & $b_0\cdot b_i$ & Notes \\
    \hline 
    $\mathfrak{g}$ & {\bf Adj} & 0 & 0 & \\ 
    \hline
    $\mathfrak{su}_N$ & $(2N)\times {\bf N}$ & $-2$ & 0 &\\
    $\mathfrak{su}_N$ & $(N-8)\times {\bf N}\oplus{\bf \frac{N(N+1)}{2}}$ & $-1$ & $-1$ & $N\ge8$\\
    $\mathfrak{su}_N$ & $(N+8)\times {\bf N}\oplus{\bf \frac{N(N-1)}{2}}$ & $-1$ & 1 &\\
    $\mathfrak{su}_N$ & $16\times {\bf N}\oplus2\times{\bf \frac{N(N-1)}{2}}$ & $0$ & 2 &\\
    $\mathfrak{su}_N$ & ${\bf \frac{N(N-1)}{2}}\oplus{\bf \frac{N(N+1)}{2}}$ & $0$ & 0 &\\
    \hline
    $\mathfrak{su}_6$ & $15\times {\bf 6}\oplus\frac{1}{2} {\bf 20}$ & $-1$ & 1 &\\
    $\mathfrak{su}_6$ & $17\times {\bf 6}\oplus{\bf 15}\oplus\frac{1}{2} {\bf 20}$ & $0$ & $2$ & \\
    $\mathfrak{su}_6$ & $18\times {\bf 6}\oplus{\bf 20}$ & 0 & 2 & \\
    $\mathfrak{su}_6$ & ${\bf 6}\oplus\frac{1}{2} {\bf 20}\oplus{\bf 21}$ & $0$ & 0 &\\
    \hline
    $\mathfrak{so}_N$ & $(N-8)\times {\bf N}$ & $-4$ & $-2$ & $N\ge 8$ \\
    $\mathfrak{so}_N$ & $(N-7)\times {\bf N}\oplus(2^{\lfloor \frac{10-N}{2}\rfloor})\times{\bf 2^{\lfloor\frac{N-1}{2}\rfloor}}$ & $-3$ & $-1$ & $12\ge N\ge 7$ \\
    $\mathfrak{so}_N$ & $(N-6)\times {\bf N}\oplus(2\times 2^{\lfloor \frac{10-N}{2}\rfloor})\times{\bf 2^{\lfloor\frac{N-1}{2}\rfloor}}$ & $-2$ & $0$ & $13\ge N\ge 6$ \\
    $\mathfrak{so}_N$ & $(N-5)\times {\bf N}\oplus(3\times 2^{\lfloor \frac{10-N}{2}\rfloor})\times{\bf 2^{\lfloor\frac{N-1}{2}\rfloor}}$ & $-1$ & $1$ & $12\ge N\ge 5$ \\
    $\mathfrak{so}_N$ & $(N-4)\times {\bf N}\oplus(4\times 2^{\lfloor \frac{10-N}{2}\rfloor})\times{\bf 2^{\lfloor\frac{N-1}{2}\rfloor}}$ & $0$ & $2$ & $14\ge N\ge 4$ \\
    \hline
    $\mathfrak{sp}_N$ & $(2N+8)\times {\bf 2N}$ & $-1$ & 1 &\\
    $\mathfrak{sp}_N$ & $16\times {\bf 2N}\oplus{\bf (N-1)(2N+1)}$ & $0$ & 2 &\\
    $\mathfrak{sp}_3$ & $\frac{35}{2} {\bf 6}\oplus \frac{1}{2} {\bf 14}' $ & $0$ & 2 &\\
    \hline
    $\mathfrak{e}_8$ &  & $-12$ & 10 &\\
    $\mathfrak{e}_7$ & $\frac{k}{2}\times {\bf 56}$ & $k-8$ & $k-6$ & $k\le 8$\\
    $\mathfrak{e}_6$ & $k\times {\bf 27}$ & $k-6$ & $k-4$ & $k\le 6$\\
    $\mathfrak{f}_4$ & $k\times {\bf 26}$ & $k-5$ & $k-3$ & $k\le 5$\\
    $\mathfrak{g}_2$ & $(3k+1)\times {\bf 7}$ & $k-3$ & $k-1$ & $k\le 3$\\
    \hline
\end{tabular}
\caption{Gauge algebras $\mathfrak{g}_i$ and types of charged hypermultiplets $H_i$ supported on $b_i$ with $b_i^2\le0$.}\label{tb:generators}
\end{table}

The proofs in this paper rely on two key results established in~\cite{Kim:2024eoa}. The first is the mutual non-negativity of distinct primitive BPS generators,
\begin{equation}
 C_i\cdot C_j\ge0,
 \qquad
 C_i\neq C_j.
 \label{prop:mutual-nonnegativity}
\end{equation}
When both generators support gauge algebras, this follows directly from the mixed gauge anomaly cancellation condition~\eqref{eq:mixed-gauge-anomaly}. The remaining cases are treated using the current algebra of the E-string, the Weyl symmetry of the M-string, and positivity of BPS-string tensions near infinite distance limits~\cite{Kim:2019vuc,Haghighat:2013gba,Lee:2019xtm,Kim:2024eoa}.

The second ingredient is the effectiveness of the gravitational anomaly charge $b_0$. As shown in~\cite{Kim:2024eoa}, supersymmetric compactification together with the strong Cobordism Conjecture \cite{McNamara:2019rup} implies that a positive integer multiple of $b_0$ is carried by a physical BPS string. Additional supporting arguments can be found in~\cite{Cheung:2016wjt,Hamada:2018dde,Kim:2024eoa} and the references therein. Equivalently, $b_0$ belongs to the real BPS cone and therefore admits a non-negative decomposition,
\begin{equation}
 b_0=\sum_i\alpha_iC_i,
 \qquad
 \alpha_i\ge0,
 \label{eq:b0-effective}
\end{equation}
where $C_i$ are primitive BPS generators. It is important to emphasize that Eq.~\eqref{eq:b0-effective} does not assume that the BPS cone itself is finitely generated; it only asserts the existence of a finite populated presentation for the single charge $b_0$. For readers familiar with F-theory, Eq.~\eqref{eq:b0-effective} is naturally interpreted as the EFT counterpart of the effectiveness of the anti-canonical divisor~\cite{Morrison:1996na,Morrison:1996pp,Morrison:2012np},

\begin{equation}
-K\ \text{effective}
\qquad\Longleftrightarrow\qquad
b_0\ \text{admits a finite populated nonnegative presentation},
\label{eq:geometric-eft-dictionary}
\end{equation}
although no geometric realization will be assumed in the present work.

The remainder of this paper shows that the two properties
\begin{equation}
b_0\in\Cone\{C_i\},
\qquad
C_i\cdot C_j\ge0
\quad(C_i\neq C_j),
\label{eq:precise-problem-inputs}
\end{equation}
together with the classification of BPS generators, are already sufficient to derive non-trivial universal bounds on the intersection numbers of primitive BPS generators. The key observation is that these two simple physical inputs naturally lead to a lattice analogue of the Zariski decomposition of $b_0$, from which the desired intersection bounds follow.

\section{Lattice Zariski decomposition of the gravitational charge}
\label{sec:zariski}

The familiar Zariski decomposition separates an effective class into a nef part and an effective negative part \cite{Zariski:1962}.  The same structure follows here directly from the BPS string spectrum, without assuming an F-theory base.  Starting from the two EFT inputs in \eqref{eq:precise-problem-inputs}, we will show that
\begin{equation}
 b_0=P+N,
 \qquad
 P\cdot C_i\geq0\quad({\rm for \ all \ generators\ } C_i),
 \qquad
 P\cdot N=0,
 \label{eq:zariski-summary}
\end{equation}
where $N$ is a positive combination of primitive BPS generators with a negative-definite intersection matrix.  We refer to this as the \emph{lattice Zariski decomposition} of $b_0$.  Its physical meaning is that $N$ selects the negative tensor sector that can become tensionless simultaneously, while $P$ is the residual gravitational charge seen non-negatively by every populated BPS string. Throughout this section and below, tensor charges $\{C_{i}\}$ denote the set of BPS generators. We also refer to a generator satisfying  $C_i^2=-n$ and $b_0\cdot C_i=2-n$ a ($-n$)-generator.

\subsection{Decomposing the gravitational anomaly charge}
\label{subsec:decomposition-proof}

\paragraph{Decomposition of $b_0$.}
\label{thm:decomposition}

The decomposition is obtained by a simple minimization procedure. Choose an arbitrary point $J$ in the interior of the tensor branch, where every BPS string has strictly positive tension, and fix one finite populated presentation
\begin{equation}
    b_0=\sum_{i=1}^{r}\beta_i C_i,
    \qquad \beta_i\geq0.
\end{equation}
For coefficients $x_i$ in the compact box $0\leq x_i\leq\beta_i$, define
\begin{equation}
    N(x)=\sum_{i=1}^{r}x_iC_i,
    \qquad
    P(x)=b_0-N(x).
\end{equation}
We call $x$ admissible when $P(x)\cdot C\geq0$ for every primitive BPS generator $C$. The admissible set is nonempty, since $x_i=\beta_i$ gives the trivial decomposition $P=0$ and $N=b_0$. It is also compact: it is the intersection of the compact box $\prod_{i=1}^{r}[0,\beta_i]$ with the closed half-spaces defined by the inequalities $P(x)\cdot C\geq0$. Since $J\cdot N(x)$ is continuous, it attains a minimum on this admissible set. We choose such a minimizing decomposition $b_0=P+N$ and refer to the generators appearing in $N$ as the \emph{support} of $N$.

The same argument yields a stronger property. Let $N'\subseteq N$ be any nonzero effective subcombination of the support. If $N'$ intersected every generator in its support non-negatively, then $N'$ could be removed from $N$ and added to $P$ without violating any of the positivity conditions, again producing an admissible decomposition with a smaller value of $J\cdot N$. Hence this is impossible. Consequently, every nonzero effective subcombination of the support has negative intersection with at least one of its constituent generators. This property will be the key ingredient in proving that the intersection matrix of the support is negative definite.

\paragraph{Negative-definite sector.}
The property established above is precisely what is needed to prove that the intersection matrix, or the Gram matrix, on the support of $N$ is negative definite. A standard result in linear algebra states that if a collection of charges has pairwise non-negative mutual intersections, and every nonzero effective subcombination has negative intersection with at least one of its constituent generators, then the corresponding intersection matrix is negative definite~\cite{BermanPlemmons:1994}.\footnote{The basic idea is as follows. If a linear combination has non-negative norm, then, because all cross intersections are non-negative, one may assume without loss of generality that its coefficients are non-negative. A non-negative combination maximizing the norm necessarily intersects every generator in its support with the same sign, contradicting the property established above unless the combination vanishes.}

This has two immediate consequences. First, every generator in the support of $N$ has negative self-intersection. Thus the minimization automatically selects the negative sector of the string charge lattice, namely the strings that can simultaneously become tensionless in a SCFT limit~\cite{Heckman:2015bfa,Bhardwaj:2015xxa,Bhardwaj:2015oru,Bhardwaj:2019hhd,Kim:2024eoa}. Second, since a negative-definite subspace of a lattice with signature $(1,T)$ has dimension at most $T$, the support of $N$ contains at most $T$ linearly independent generators, regardless of how large the BPS spectrum is.

Finally, because $P$ intersects every primitive generator non-negatively while $b_0$ is an effective combination of generators, we have
\begin{align}
    P\cdot P=P\cdot b_0\ge0.
\end{align}
Thus $P$ is necessarily a time-like or null charge. This completes the Zariski decomposition of $b_0$ from EFT perspective.

Combining this with the orthogonality relation $P\cdot N=0$, we obtain
\begin{equation}
  b_0\cdot b_0 = P\cdot P + N\cdot N,
  \label{eq:pythagoras}
\end{equation}
which decomposes the norm of the gravitational charge $b_0^2=9-T$~\cite{Green:1984bx,Sagnotti:1992qw,Erler:1993zy} into a non-negative contribution carried by $P$ and a negative contribution coming from the sector $N$.

\paragraph{Uniqueness.}
We can prove that the minimal decomposition of $b_0$ is unique, so it admits a canonical decomposition. It
depends only on $b_0$ and the list of BPS generators, not on how $b_0$ was
presented as an effective combination.  The reason is the same interplay
of positivity structure.  If two splittings $P+N$ and $P'+N'$ both satisfied the conditions of~\eqref{eq:zariski-summary}, their difference
$z=P-P'=N'-N$ would have non-negative square, since
$z\cdot z=P\cdot N'+P'\cdot N$ and each term is non-negative. However, $z$ is
also a difference of two negative-definite sectors, and because distinct generators never pair negatively, such a difference can have non-negative square only if $z$ itself vanishes. The two splittings therefore coincide.

\subsection{Attractor interpretation}
The lattice Zariski decomposition admits a natural interpretation in terms of the attractor mechanism for six-dimensional black strings~\cite{Ferrara:1995ih,Ferrara:1996dd,Haghighat:2015ega,HetLam:2018yba}. Motivated by Denef's study of split attractor flows and attractor flow trees in 4d $\mathcal{N}=2$ theories~\cite{Denef:2000nbc}, where attractor flows make sense for BPS charges which are not single-centered black holes, we can study the attractor flow for BPS strings that are not necessarily black strings. A BPS string of charge $Q$ drives the tensor moduli toward the point that extremizes its central charge. When the unconstrained attractor point lies outside the physical tensor branch because $Q$ pairs negatively with effective strings, the attractor flow instead reaches the boundary of the tensor branch. 

The lattice Zariski decomposition identifies this endpoint explicitly. When $P\cdot P>0$, the attractor point is 
\begin{align}
  J_*=\frac{P}{\sqrt{P\cdot P}},
\end{align}
which lies on the closure of the tensor branch. At this point every generator in the support of $N$ becomes tensionless,
\begin{align}
    J_*\cdot C_i=0,
\end{align}
while all other primitive BPS strings retain non-negative tension. Thus the attractor flow singles out precisely the negative-definite sector encoded by $N$, corresponding to the superconformal degeneration in which these strings become simultaneously tensionless. 

The decomposition $b_0=P+N$ then can be viewed as the constrained attractor decomposition of the gravitational anomaly charge. The timelike component $P$ determines the attractor direction, while the negative-definite component $N$ is supported on the constraints that are saturated at the boundary. Also, since a finite-distance degeneration generated by a negative-definite set of string charges defines a superconformal sector~\cite{Heckman:2015bfa,Bhardwaj:2015xxa,Bhardwaj:2019hhd}, the strings in the support of $N$ describe the corresponding tensionless strings in the SCFT sector.

\subsection{Generators in the support of $N$}
\label{sec:zariski-support}

We now determine which primitive BPS generators are necessarily contained
in the support of the negative part $N$. Recall that
\begin{equation}
 b_0=P+N,
 \qquad
 N=\sum_{i\in S}\alpha_i C_i,
 \qquad
 \alpha_i>0,
\end{equation}
where $P$ intersects every BPS generator non-negatively and
$S=\operatorname{Supp}(N)$. Since distinct generators also intersect non-negatively, these positivity properties immediately imply the criterion
\begin{align}\label{eq:N-support-generator}
    b_0\cdot C<0 \qquad\Longrightarrow\qquad  C\in S \ .
\end{align}
In physical terms, every string whose charge pairs negatively with the gravitational charge $b_0$necessarily belongs to the tensionless sector selected by the attractor.

To prove \eqref{eq:N-support-generator}, suppose instead that $C\notin S$. Since every generator appearing in $N$ is distinct from $C$, mutual non-negativity together with the positive expansion of $N$ implies $N\cdot C\ge0$. Combining this with $P\cdot C\ge0$, we obtain
\begin{align}
    b_0\cdot C=P\cdot C+N\cdot C\geq0,
 \label{eq:minus-n-support-contradiction}
\end{align}
which contradicts the assumption $b_0\cdot C<0$. This establishes our statement in \eqref{eq:N-support-generator}. 

An immediate consequence is that every $(-n)$-generator with $n\ge3$ necessarily belongs to $S$, since $b_0\cdot C=2-n<0$ from Table~\ref{tb:generators}. The same argument also shows that a generator satisfying $C^2=-1$ and $b_0\cdot C=-1$ must belong to $S$.

For a $(-2)$-generator, on the other hand, the zero intersection with $b_0$ leads to a different conclusion. Since $b_0\cdot C_{-2}=0$, the criterion \eqref{eq:N-support-generator} alone does not determine whether $C_{-2}$ belongs to $S$.

Suppose first that $C_{-2}$ intersects a generator $D\in S$. If $C_{-2}\notin S$, then
\begin{align}
    N\cdot C_{-2} = \sum_{i\in S}\alpha_i(C_i\cdot C_{-2}) >0,
\end{align}
because every coefficient $\alpha_i$ is positive and the contribution of $D$ is strictly positive. Together with $P\cdot C_{-2}\ge0$, this gives
\begin{align}
    b_0\cdot C_{-2}=P\cdot C_{-2}+N\cdot C_{-2}>0,
\end{align}
contradicting $b_0\cdot C_{-2}=0$. Therefore every $(-2)$-generator intersecting a generator already contained in $S$ must itself belong to $S$.
Applying this argument recursively, the support $S$ necessarily contains the entire connected component of the $(-2)$ intersection graph attached to any generator with $b_0\cdot C<0$. In particular, every $(-2)$-generator connected to a $(-3)$- or $(-4)$-generator is contained in $S$.

The remaining $(-2)$-generators are those contained in $S$ but isolated from all other generators in $S$, and those lying completely outside $S$. We refer to these as {\it isolated} $(-2)$-{\it generators}. For the latter group, we can also show that $P\cdot C_{-2}=N\cdot C_{-2}=0$, which follows directly from $b_0\cdot C_{-2}=0$.

We may naturally expect that there can not be any isolated $(-2)$-generators lying inside $S$. For F-theory examples, we have the following argument: the elliptic Calabi--Yau fibration over the base $B$ obeys a canonical bundle formula~\cite{Fujita:1986,Vafa:1996xn,Morrison:1996na,Morrison:1996pp}, which in physics language corresponds to Type IIB supergravity equation of motion: $b_0=c_1(B)=Ric_B$~\cite{Sadov:1996zm} has a smooth part $\partial_i \tau\partial_ {\bar i}\bar\tau$ from variation of 10d axio-dilaton, and also a delta-function part supported on the loci of 7-branes, whose coefficient is the deficit angle (normalized by $\frac{1}{2\pi}$) for the corresponding 7-brane~\cite{Greene:1989ya,Weigand:2018rez} (which is 0 for (p,q) 7-branes and $\frac{1}{2}$ for O7 brane, and take values $\frac{1}{6},\frac{5}{6},\frac{1}{4},\frac{3}{4},\frac{1}{3},\frac{2}{3}$ for Kodaira type $II,II^*,III,III^*,IV,IV^*$ singularities.) 
This is one admissible decomposition of $b_0$, but might not be the optimal one by our minimum definition of the Zariski decomposition, we see that the coefficient of generators $C_i$ in $N$ is at most its deficit angle. but isolated $(-2)$ curves cannot support a 7-brane with nonzero deficit angle, as all of them come from orbifold constructions that decrease the self-intersection from $-2$. If the $(-2)$ curve is not isolated, there might be an opposite contribution from its intersection with other orbifold locus, hence these are not excluded from $N$, while isolated $(-2)$ curves are excluded from $N$.

Finally, ordinary $(-1)$-generators, which we denoted by $e$, satisfying $e^2=-1$ and $b_0\cdot e=1$ are not determined by the positivity criterion. They may or may not belong to the support, depending on the complete negative-definite intersection matrix of $S$. We note in F-theory that, by the previous argument, they cannot belong to the support.

\section{Bounds on the intersection numbers}
\label{sec:bounds}
\label{sec:masterbound}

The geometry of the tensor branch is completely determined by a finite amount of lattice data: the intersection matrix $\Omega$ of the tensor multiplets together with the anomaly vectors $b_0$ and $b_i$. Equivalently, this information can be encoded in the primitive BPS generators of the string charge cone. Their self-intersections determine the types of generators that can appear, while their mutual intersections determine the intersection graph of the cone and hence the possible patterns of simultaneous tensionless strings. Once the allowed generator types have been classified, the remaining discrete data are precisely the intersection numbers among the primitive generators. Determining the allowed values of these intersection numbers is therefore an essential step toward reconstructing the tensor branch, and ultimately the complete tensor sector of 6d (1,0) supergravity~\cite{Kim:2024eoa,Hamada:2026zta}.

In this section, we first review the known constraints on these intersection numbers arising from gauge anomaly cancellation and from the unitarity of the current algebras on the worldsheet CFTs on BPS strings. We then show that the lattice Zariski decomposition developed in the previous section provides an additional universal constraint on the intersections among BPS generators. Combining this new constraint with the previously known ones, we derive universal bounds on all remaining intersection numbers that were not fixed by anomaly cancellation or worldsheet unitarity alone.

\subsection{Known constraints and remaining open cases}
\label{sec:known-bounds}

Several mechanisms are already known to constrain tensor intersection numbers. First, when two generators both support gauge algebras, their mutual intersection is fixed by the mixed gauge-anomaly equation~\eqref{eq:mixed-gauge-anomaly}.  The intersection number is then determined by the spectrum of matter charged under both gauge factors. 

Second, suppose one generator supports a gauge algebra $\mathfrak g$ with anomaly vector $b_{\mathfrak g}$, while the other is either the E-string charge $e$ or an M-string charge $C_{-2}$. The intersection number determines the level $k_{\mathfrak g}=Q\cdot b_{\mathfrak g}$ of the $\mathfrak g$ current algebra realized on the 2d worldsheet CFT on BPS strings with charge $Q=e,C_{-2}$. Worldsheet unitarity then constrains the allowed values of $k_{\mathfrak g}$ as discussed in \cite{Kim:2019vuc,Lee:2019skh}.

For the E-string, the induced current algebra is embedded in the level-one $E_8$ current algebra in the left-moving sector. Consequently, the embedding must be compatible with the $(E_8)_1$ current algebra, and the corresponding Sugawara central charge cannot exceed the total left-moving central charge of the E-string worldsheet theory~\cite{DiFrancesco:1997nk,Minahan:1998vr,Kim:2019vuc},
\begin{align}\label{eq:central-charge-embedding}
    c_{\mathfrak g} = \frac{k_{\mathfrak g}\,\dim\mathfrak g}{k_{\mathfrak g}+h^\vee_{\mathfrak g}} \le c_L=8\, ,
\end{align}
where $h^\vee_{\mathfrak g}$ is the dual Coxeter number of $\mathfrak g$. Similarly, for the M-string, the induced current algebra must be embedded in the level-one $SU(2)$ current algebra, leading to analogous constraints on the intersection number.

For generators with self-intersection $-n$ $(n\ge5)$, the above constraints completely determine the allowed intersections. Mixed gauge anomaly cancellation forbids intersections with any other generator supporting a gauge algebra. Intersections with M-strings are likewise excluded by the requirement that the induced current algebra be embedded into the level-one $SU(2)$ current algebra on the M-string worldsheet theory. Consequently, a $(-n)$-generator with $n\ge5$ can intersect only the E-string.
Moreover, the intersection with the E-string is itself constrained by the requirement that the induced current algebra be embedded into the level-one $E_8$ current algebra of the E-string. This leaves the bound~\cite{Heckman:2015bfa,Hamada:2026zta}:
\begin{align}
    C_{-n}\cdot e \le 1, \qquad n\ge5.
\end{align}

The situation is similarly rigid for a $(-4)$-generator. It may intersect a $(-1)$-generator with maximal intersection number two. When the $(-1)$-generator supports a gauge algebra, this follows directly from anomaly cancellation, while for the E-string the same bound follows from the $E_8$ current algebra embedding. A $(-4)$-generator may also intersect a $(-2)$-generator, but only when the latter supports a gauge algebra, in which case anomaly cancellation fixes the intersection number to be two. So we have~\cite{Bhardwaj:2018jgp,Bhardwaj:2019hhd,Hamada:2026zta}
\begin{align}
    C_{-4}\cdot e \le 2 \quad {\rm and} \quad C_{-4}\cdot C_{-2} = 0,2 \ .
\end{align}

The same reasoning applies to a generator $\hat C$ satisfying $\hat C^2=-1$ and $b_0\cdot\hat C=-1$. Such a generator can appear only with frozen singularity in string theory~\cite{Bhardwaj:2018jgp,Bhardwaj:2019hhd,MorrisonSung:2023}. It may intersect only $(-1)$- or $(-2)$-generators. Moreover, anomaly cancellation together with the worldsheet current algebra constraints implies~\cite{Bhardwaj:2018jgp,Bhardwaj:2019hhd,Hamada:2026zta}
\begin{align}
    \hat C\cdot e\le1\quad {\rm and} \quad \hat C\cdot C_{-2}\le1\ .
\end{align}

The first genuinely unconstrained case appears for a $(-3)$-generator. A $(-3)$-generator can intersect only $(-2)$- or $(-1)$-generators. If the $(-2)$-generator supports a gauge algebra, anomaly cancellation fixes the intersection number to one. Likewise, if the $(-1)$-generator supports a gauge algebra, the intersection number is also fixed to one~\cite{Heckman:2015bfa,Bhardwaj:2019hhd,Hamada:2026zta}. The only remaining possibility is its intersection with an E-string, for which no universal bound was previously known. One of the main results of the present work is to derive such a bound directly from the lattice Zariski decomposition.

The remaining open problem concerns intersections involving $(-2)$-generators. When both generators support gauge algebras, anomaly cancellation again fixes their intersection number to one or two~\cite{Bhardwaj:2015xxa,Bhardwaj:2019hhd,Hamada:2026zta}. However, this argument becomes ineffective for small gauge algebras such as $\mathfrak{su}(2)$ and $\mathfrak{su}(3)$, as well as for generators without gauge algebras. Consequently, no universal bounds were previously known for intersections among $(-2)$-generators or between a $(-2)$-generator and an E-string. The remainder of this section is devoted to deriving these bounds.

Finally, intersections among the $(-1)$-generators themselves are of a different character. It turns out that, once the intersection structure of all other generators is fixed, the mutual intersections of the $(-1)$-charges are determined systematically by the structure of the BPS string cone.  This reconstruction is treated in a companion paper, and we do not discuss it further here.

\subsection{Bounds from Zariski decomposition}
\label{sec:zariski-intersection-bounds}

We now use the lattice Zariski decomposition to derive universal bound on the intersection numbers that remained open in the previous subsection.  Our main focus in this subsection is the intersection between a $(-1)$-generator $e$ and the generators contained in the support $S=\operatorname{Supp}(N)$. Then the only remaining case not covered by this analysis is the intersection between an isolated $(-2)$-generator and an E-string, which will be discussed separately in the next subsection. 

Suppose first that the $(-1)$-generator $e$ is contained in the support $S$. This means that the $(-1)$-string becomes tensionless together with every other strings in $S$ at the attractor point. Therefore, the generators in $S$ form 6d SCFT sector, and therefore their mutual intersections must satisfy the intersection rules of tensor multiplets in  6d SCFTs. Hence, from the classification of 6d SCFTs~\cite{Heckman:2015bfa,Bhardwaj:2015xxa,Bhardwaj:2019hhd}, one immediately obtains
\begin{align}
    C_{-3}\cdot e\le1 \quad {\rm and} \quad C_{-2}\cdot e\le1, \qquad C_{-3},\,C_{-2},\,e\in S.
    \label{eq:internal-bound}
\end{align}

We now consider the case where the $(-1)$-generator is not contained in the support $S$. Pairing the lattice Zariski decomposition with $e$ gives
\begin{align}
    b_0\cdot e = P\cdot e+N\cdot e = P\cdot e+\sum_{i\in S}\alpha_i(C_i\cdot e)\, .
\label{eq:external-master-relation}
\end{align}
Every term on the right-hand side is non-negative. It therefore follows that
\begin{align}
    \alpha_i(C_i\cdot e)\le1\ , \qquad \forall\,C_i\in S,
\end{align}
which immediately implies
\begin{align}
    C_i\cdot e \le \left\lfloor\frac{1}{\alpha_i}\right\rfloor.
\label{eq:external-master-bound}
\end{align}
Thus the coefficient of each generator appearing in $N$ directly determines an upper bound on its intersection with an external $(-1)$-generator not in $S$.

Applying this result to the one-node $(-3)$-sector gives the desired bound for an isolated $(-3)$-generator. We then show that coupling the $(-3)$-generator to a larger negative sector can only increase its coefficient in $N$ and therefore strengthen, rather than weaken, the intersection bound. The same reasoning determines the bounds for the $(-2)$-generators contained in non-Higgsable clusters.

\paragraph{Isolated $(-3)$-generator.}
We begin with an isolated $(-3)$-generator $C_{-3}$. Since
\begin{align}
    C_{-3}^2=-3\, ,\qquad b_0\cdot C_{-3}=-1\, ,
\end{align}
it necessarily belongs to the support $S$. The local contribution of $C_{-3}$ to the negative part can therefore be written as
\begin{align}
    N=\alpha_{-3}C_{-3}+N'\, ,
\end{align}
where $N'$ consists of the remaining generators in the support.

Since the $(-3)$-generator is isolated, every generator appearing in $N'$ intersects $C_{-3}$ trivially. The orthogonality condition $P\cdot C_{-3}=0$ then gives
\begin{align}
    0
    &=(b_0-N)\cdot C_{-3}\nonumber\\
    &=(b_0-\alpha_{-3}C_{-3})\cdot C_{-3}\nonumber\\
    &=-1+3\alpha_{-3}\, ,
\end{align}
which immediately yields
\begin{align}
    \alpha_{-3}=\frac13\, .
\end{align}
Substituting this into Eq.~\eqref{eq:external-master-bound}, we obtain
\begin{align}
    \boxed{C_{-3}\cdot e\le3} \ .
    \label{eq:isolated-minus-three-bound}
\end{align}
This bound follows entirely from the lattice Zariski decomposition together with the unit pairing $b_0\cdot e=1$. Unlike the previously known bounds reviewed in the last subsection, it does not rely on gauge anomaly cancellation or on worldsheet current algebra constraints.

\paragraph{Generalization to larger block.}
The derivation above treated the $(-3)$-generator as an isolated negative block. In a general theory, however, the support of $N$ may contain additional generators and the $(-3)$-generator is no longer isolated. One must therefore show that enlarging the support cannot decrease the coefficient $\alpha_{-3}$, since otherwise the intersection bound could become weaker.

To see this, divide the support into the block $B$ containing the $(-3)$-generator and the remaining generators $R$. The orthogonality conditions $P\cdot C_i=0$ for $i\in B$ become
\begin{align}
  \sum_{j\in B}\alpha_j\,C_j\cdot C_i
  \;=\;
  b_0\cdot C_i \;-\; \sum_{k\in R}\alpha_k\,C_k\cdot C_i ,
  \qquad i\in B\ .
\end{align}
Since $\alpha_k\ge0$ and distinct generators intersect
non-negatively, the second term on the right-hand side is
non-positive. Let us now define
\begin{align}
     M_{ij}=-C_i\cdot C_j\, , \qquad q_i=-b_0\cdot C_i\quad {\rm for} \quad i\in B \, .
\label{eq:M-matrix-definitions}
\end{align}
Solving the block equations amounts to multiplying by $M_{BB}^{-1}$, where $M_{BB}$ is a symmetric positive-definite matrix with positive diagonal entries and non-positive off-diagonal entries, namely a non-singular $M$-matrix (or equivalently a Stieltjes matrix). A standard property of such matrices is that their inverse is entrywise non-negative~\cite{BermanPlemmons:1994}. Therefore,
\begin{align}
    \alpha_B = M_{BB}^{-1} \bigl(q_B-M_{BR}\alpha_R\bigr) \ge M_{BB}^{-1}q_B = \alpha_B^{\rm block}\,,
\end{align}
which proves the monotonicity
\begin{align}
    \alpha_i\ge\alpha_i^{\rm block}\, , \qquad i\in B.
\label{eq:M-monotonicity}
\end{align}

Thus embedding a local negative block into a larger support can only increase its Zariski coefficients $\alpha_i$. Consequently, every intersection bound obtained from an isolated block is universal. It remains valid in an arbitrary populated spectrum and can only become stronger. Then, together with the previous results, this establishes the universal intersection bound $C_{-3}\cdot e\le 3$.

\paragraph{$(-2)$-generators in the support $S$.}
\label{sec:blocks}

A $(-2)$-generator attached to another generator already contained in $S$ must itself belong to $S$, as shown in Section~\ref{sec:zariski-support}. Such a $(-2)$-generator must either belong to a non-Higgsable cluster or lie in a connected component attached to a $(-4)$-generator. In the latter case, the $(-2)$-generator necessarily supports a gauge algebra at least as large as $SU(6)$, and thus its intersection with a $(-1)$-generator is already bounded by one by the worldsheet current algebra constraint~\cite{Morrison:2012np,Hamada:2026zta}. Therefore, the only remaining case is a $(-2)$-generator is an element of a NHC (possibly with enhanced gauge symmetry).

There are precisely three NHCs containing $(-2)$-generators, namely the chains $(-3,-2)$,
$(-3,-2,-2)$, and $(-2,-3,-2)$ classified in~\cite{Morrison:2012np,Heckman:2015bfa,Bhardwaj:2019hhd}. For each of these clusters, the Zariski coefficients are determined by the coupled orthogonality conditions. As an example, consider the $(-3,-2)$ block. Writing
\begin{equation}
  N=\alpha_1C_{-3}+\alpha_2C_{-2} + \cdots \,,
\end{equation}
the orthogonality equations become
\begin{equation}
  -3\alpha_1+\alpha_2=-1,
  \qquad
  \alpha_1-2\alpha_2=0
  \qquad\Longrightarrow\qquad
  (\alpha_1,\alpha_2)=\bigl(\tfrac25,\tfrac15\bigr) \ .
\end{equation}
Therefore, we compute the bounds on the intersection numbers as
\begin{equation}
  C_{-3}\cdot e\leq2,
  \qquad
  C_{-2}\cdot e\leq5.
\end{equation}
Repeating the same calculation for the remaining NHCs yields the following coefficients and intersection bounds:
\begin{center}
\begin{tabular}{lll}
\hline
NHC & $\alpha_i$ & bounds on $C\cdot e$ \\
\hline
$(-3)$ & $\tfrac13$ & $3$ \\
$(-3,-2)$ & $\bigl(\tfrac25,\ \tfrac15\bigr)$ & $(2,\ 5)$ \\
$(-3,-2,-2)$ & $\bigl(\tfrac37,\ \tfrac27,\ \tfrac17\bigr)$ & $(2,\ 3,\ 7)$ \\
$(-2,-3,-2)$ & $\bigl(\tfrac14,\ \tfrac12,\ \tfrac14\bigr)$ & $(4,\ 2,\ 4)$ \\
\hline
\end{tabular}
\end{center}
The monotonicity result in~\eqref{eq:M-monotonicity} shows that embedding any of these local blocks into a larger negative subsystem can only increase their Zariski coefficients. Consequently, the bounds obtained above remain valid in any general cases. In particular, every $(-2)$-generator belonging to the support satisfies the universal bound
\begin{align}
    \boxed{C_{-2}\cdot e \le 7 \quad {\rm for} \quad C_{-2} \in S} \ .
\end{align}

\paragraph{Comparison with F-theory bounds.}
\label{subsec:f-theory-comparison}

It is useful to compare our EFT bounds with the intersection patterns realized in F-theory. Morrison and Taylor classified the allowed ways in which a $(-1)$ curve can meet a non-Higgsable cluster. The relevant configurations are summarized in Table~3 of Ref.~\cite{Morrison:2012np}. Their analysis uses additional geometric input, including the vanishing orders of the Weierstrass coefficients and the exclusion of non-minimal singularities, neither of which enters our EFT derivation. Comparing the two results therefore illustrates how much of the geometric constraints already follows from low-energy consistency alone and where genuinely geometric input becomes necessary. The comparison is summarized in Table~\ref{tab:eft-f-theory}.

\begin{table}[t]
\centering
\small
\renewcommand{\arraystretch}{1.2}
\begin{tabular}{>{\raggedright\arraybackslash}p{0.34\textwidth}
                >{\centering\arraybackslash}p{0.16\textwidth}
                >{\centering\arraybackslash}p{0.18\textwidth}
                >{\raggedright\arraybackslash}p{0.19\textwidth}}
\toprule
Charges & EFT bound & F-theory bound & Comparison \\
\midrule
isolated $(-3)$ & $C\cdot e\leq3$ & $C\cdot e\leq2$ & EFT weaker by one \\
$(-3)$ in $(-3,-2)$ & $C\cdot e\leq2$ & $C\cdot e\leq2$ & agreement \\
central $(-3)$ in $(-2,-3,-2)$ & $C\cdot e\leq2$ & $C\cdot e\leq2$ & agreement \\
$(-3)$ in $(-3,-2,-2)$ & $C\cdot e\leq2$ & $C\cdot e\leq2$ & agreement \\
$(-2)$ in $(-3,-2)$ & $C\cdot e\leq5$ & $C\cdot e\leq2$ & EFT weaker \\
end $(-2)$ in $(-2,-3,-2)$ & $C\cdot e\leq4$ & $C\cdot e\leq2$ & EFT weaker \\
first/second $(-2)$ in $(-3,-2,-2)$ & $C\cdot e\leq3,7$ & $C\cdot e\leq2$ & EFT weaker \\
\bottomrule
\end{tabular}
\caption{EFT intersection bounds for NHCs compared with the geometric bounds of Ref.~\cite{Morrison:2012np}.}
\label{tab:eft-f-theory}
\end{table}

The comparison reveals several interesting features. First, the EFT bounds reproduce the geometric bounds exactly for all $(-n)$-generators with $n\ge4$, as discussed in Section~\ref{sec:known-bounds}. They also agree with the geometric bounds for several $(-3)$ components and differ by only one in the case of an isolated $(-3)$-generator. By contrast, the EFT bounds are weaker for some $(-2)$ components contained in $S$, although the discrepancy is rather small. This discrepancy reflects additional UV information encoded in the F-theory geometry that is not captured by the low-energy EFT arguments used here. Filling this gap, possibly through more general quantum gravity consistency conditions, would be an interesting direction for future work.

\paragraph{Isolated $(-2)$-generators.}
As discussed so far, the lattice Zariski decomposition allows us to determine the intersection bounds for every generator contained in the support of the negative part $N$. The only remaining case is an isolated $(-2)$-generator supporting a gauge algebra. Since $b_0\cdot C_{-2}=0$, such a generator need not appear in the support $S$, and its Zariski coefficient $\alpha_i$ may vanish. Consequently, the Zariski coefficient method does not provide any constraint on its intersection with an E-string.

To study this remaining case, we instead use the unitarity constraint on the 2d worldsheet theory of the E-string. Let $C_{-2}$ support a gauge algebra $\mathfrak g$. When the intersection number is $k$, their intersection induces a $\mathfrak g$ current algebra on the E-string worldsheet at level $k_{\mathfrak g}=k$~\cite{Kim:2019vuc,Shimizu:2016lbw}. Consistency of the worldsheet theory requires this current algebra to embed inside the left-moving $(E_8)_1$ current algebra carried by the E-string. For every simple gauge algebra other than $SU(2)$ and $SU(3)$, the central charge bound in \eqref{eq:central-charge-embedding} already gives a stronger constraint on the level and thus on the intersection number. The only remaining possibilities are therefore $\mathfrak g=SU(2)$ and $SU(3)$. Since the $SU(2)$ case gives the weakest constraint, it suffices to consider it.

The input we will use is the classification of $\mathfrak{sl}_2$ subalgebras of $\mathfrak e_8$ due to Dynkin~\cite{Dynkin:1957}. These subalgebras form finitely many conjugacy classes, each characterized by its Dynkin index, with the largest index realized by the principal $\mathfrak{sl}_2\subset\mathfrak e_8$ embedding. More generally, for a simple Lie algebra $\mathfrak g$, the Dynkin index of the principal embedding is~\cite{Panyushev:2009,Panyushev:2014}
\begin{equation}
  I_{\rm max}
  =
  \frac{1}{h^\vee}
  \sum_{m\in{\rm exp}(\mathfrak g)}
  \frac{m(m+1)(2m+1)}{3},
  \label{eq:principalindex}
\end{equation}
where the sum runs over the exponents of $\mathfrak g$.
Evaluating this equation for some Lie algebras gives
\begin{center}
\begin{tabular}{lcccccc}
\hline
$\mathfrak{g}$ & $\mathfrak{su}(3)$ & $\mathfrak{g}_2$ & $\mathfrak{f}_4$
 & $\mathfrak{e}_6$ & $\mathfrak{e}_7$ & $\mathfrak{e}_8$ \\
\hline
principal index & $4$ & $28$ & $156$ & $156$ & $399$ & $1240$ \\
\hline
\end{tabular}
\end{center}
In particular, for $\mathfrak e_8$, it gives
\begin{align}
    I_{\rm max}=1240,
\end{align}
using the exponents $1,7,11,13,17,19,23,29$ and $h^\vee(E_8)=30$~\cite{Panyushev:2009,Panyushev:2014}.

For an affine embedding $\mathfrak h\subset\mathfrak g$, the induced level of the subalgebra is multiplied by the Dynkin index of the embedding, $k_{\mathfrak h}=I_{\mathfrak h\hookrightarrow\mathfrak g}\,k_{\mathfrak g}$~\cite{DiFrancesco:1997nk}. Since the E-string carries the $(E_8)_1$ current algebra, the induced $SU(2)$ current algebra has level equal to the embedding index itself. Combining this with the maximal principal index therefore yields the universal bound
\begin{equation}
    \boxed{C_{-2}\cdot e\le1240}\,,
\end{equation}
for an isolated $(-2)$-generator. This is a universal upper bound on the induced worldsheet current level, and hence on the intersection number between an isolated $(-2)$-generator and the E-string. However, we do not claim that the principal embedding can necessarily be realized in a complete 6d supergravity theory, and we expect that the optimal bound for consistent quantum gravity theories is considerably smaller.

\subsection{Summary and finiteness up to charge-lattice duality}
\label{subsec:finiteness-up-to-duality}

\begin{table}[t]
\centering
\small
\renewcommand{\arraystretch}{1.2}
\begin{tabular}{>{\raggedright\arraybackslash}p{0.20\textwidth}
                >{\raggedright\arraybackslash}p{0.48\textwidth}
                >{\centering\arraybackslash}p{0.19\textwidth}}
\toprule
Charges & Method & Bound \\
\midrule
Isolated $(-3)$ & Zariski coefficient & $C\cdot e\leq3$ \\
$(-2)$ in NHC & Zariski coefficient + $M$-matrix monotonicity & $C\cdot e\leq3,4,5,7$ \\
Isolated $(-2)$ & $(E_8)_1$ embedding index & $C\cdot e\leq1240$ \\
\bottomrule
\end{tabular}
\caption{Coverage of the intersection bounds and the physical input responsible for each.}
\label{tab:combined-coverage}
\end{table}

We conclude this section by summarizing the mechanisms that constrain the intersection numbers between primitive BPS generators. Intersections between two generators supporting gauge algebras are fixed by the mixed gauge anomaly equation~\eqref{eq:mixed-gauge-anomaly}. Intersections between an E-string and a $(-n)$-generator with $n\ge4$, or a generator $\hat C$ satisfying $\hat C^2=b_0\cdot\hat C=-1$, are bounded by the worldsheet current algebra and the central-charge constraint \eqref{eq:central-charge-embedding}. The new results of this paper concern the remaining cases. Intersections with $(-3)$-generators and with $(-2)$-generators contained in non-Higgsable clusters are bounded by the lattice Zariski decomposition, while isolated $(-2)$-generators carrying gauge algebras are controlled by the embedding of the induced current algebra into the left-moving $(E_8)_1$ algebra on the E-string worldsheet.

These complementary mechanisms are summarized in Table~\ref{tab:combined-coverage}. Together they show that every intersection between primitive BPS generators is either fixed, bounded, or removed by charge-lattice duality. Under the two assumptions introduced in Section~\ref{subsec:bps-cone} and used in~\cite{Kim:2024eoa}, this establishes the finiteness of the tensor-intersection sector up to tensor charge-lattice duality.

The only intersections not analyzed here are those between distinct E-string generators. Nevertheless, they do not introduce an infinite family of independent tensor intersection data. Once the intersections between E-strings and all other primitive generators are fixed, the remaining E-string intersections are expected to be determined systematically by the requirement that all primitive generators satisfy the mutual non-negativity condition. A detailed analysis of this issue will be presented in a separate paper.

\section{Discussion}
\label{sec:discussion}

The central result of this paper is that the previously uncontrolled intersections between an E-string and negative tensor generators are bounded by a small set of low-energy consistency conditions.  The key ingredient is the lattice Zariski decomposition $b_0=P+N$ of the gravitational charge $b_0$, whose physical content is the attractor flow of the gravitational BPS black string~\cite{Ferrara:1995ih,Ferrara:1996dd,HetLam:2018yba}. This mechanism gives the bound $C_{-3}\cdot e\le3$ for a $(-3)$-generator, and $C_{-2}\cdot e\le7$ for a $(-2)$-generator contained in NHCs. For an isolated gauged $(-2)$-string, we prove a rather weaker bound $C_{-2}\cdot e\le1240$ using the finite set of current-algebra embeddings into the $(E_8)_1$ worldsheet algebra of the E-string~\cite{Minahan:1998vr,Kim:2014dza,Kim:2019vuc,Dynkin:1957,Panyushev:2009,Panyushev:2014}. Importantly, no string theory or F-theory realization enters the derivation at any point, and the proofs are purely from EFT arguments.

An important consequence is that every intersection between generators of BPS string charge cone is now either fixed by anomaly cancellation, bounded by the lattice Zariski decomposition, bounded by worldsheet current algebra, or removed by charge-lattice duality. Under the tensor-boundary and completeness assumptions introduced in~\cite{Kim:2024eoa}, this establishes the finiteness of the primitive tensor-intersection data up to charge-lattice duality.  We expect this result to provide an important step toward the complete EFT classification of tensor sectors in 6d supergravity initiated in~\cite{Hamada:2026zta}.

\subsection{EFT versus F-theory geometry}
\label{subsec:discussion-f-theory}

The comparison with F-theory separates universal EFT information from constraints specific to a geometric UV completion.  For the $(-3)$ components of the $(-3,-2)$, $(-2,-3,-2)$, and $(-3,-2,-2)$ clusters, the EFT calculation reproduces the F-theory maximum two.  For an isolated $(-3)$-generator, it gives three rather than two.  This near agreement is nontrivial because the EFT derivation uses neither Weierstrass vanishing orders nor the exclusion of codimension-two non-minimal singularities used in~\cite{Morrison:2012np}.

The difference is larger for some $(-2)$ components in the negative support.  Their Zariski coefficients give the bounds $3$, $4$, $5$, and $7$, while the corresponding F-theory multiplicities are at most two~\cite{Morrison:2012np}. These results are compatible: the EFT bounds apply to every populated charge lattice satisfying the assumptions of Section~\ref{subsec:bps-cone}, whereas the geometric bounds incorporate additional UV data.  The gap therefore identifies concrete low-energy configurations on which a stronger quantum gravity condition would have to act.

There is also a difference in what is being bounded.  The Zariski argument gives a componentwise pairing with one external E-string, while the geometric classification constrains the complete incidence pattern of that curve with an entire cluster~\cite{Morrison:2012np}.  Bounding such simultaneous intersections directly in EFT may require using the full relation \eqref{eq:external-master-relation}, together with anomaly cancellation and  current algebra constraints, rather than estimating one non-negative term at a time.

\subsection{AI-guided discovery process}
\label{subsec:discussion-ai}

The central idea of this work did not emerge directly from the analytic investigation of tensor intersections by authors. Rather, it arose during an AI-assisted search for an organizing principle underlying the tensor charge cone of 6d $\mathcal{N}=(1,0)$ supergravity. 

The original question was whether the intersection numbers between primitive BPS generators could be constrained using only effective field theory. More precisely, starting from the positivity of the gravitational anomaly charge,
\begin{equation}
    b_0=\sum_i\beta_iC_i, \qquad \beta_i\geq0\,,
    \label{eq:discussion-original-population}
\end{equation}
together with the mutual non-negativity condition $C_i\cdot C_j\ge0$ for distinct primitive generators, we asked whether one could derive universal bounds on the intersection numbers $C\cdot e$ without using any string theory or F-theory inputs~\cite{Kim:2024eoa}.

Early attempts focused on the intersections of individual pairs of generators, particularly the $(-2)$- and $(-3)$-generators together with the E-string. Although these investigations suggested that non-trivial bounds might exist, repeated discussions with Claude Opus 4.8 and ChatGPT 5.5 did not lead to a satisfactory EFT derivation.

A conceptual breakthrough occurred during subsequent AI-assisted exploration. Claude Fable 5 suggested that the gravitational anomaly charge itself, rather than individual generator pairs, should be the fundamental object of study, and proposed that a Zariski-type decomposition of $b_0$ might provide the missing organizing principle~\cite{Zariski:1962}. Motivated by this observation, we investigated, again with Claude Fable 5, whether such a decomposition could be formulated purely on the tensor charge lattice. This eventually led to the lattice Zariski decomposition introduced in this paper, in which the gravitational charge is decomposed into a part that intersects every primitive generator non-negatively and a complementary negative-definite part supported on a tensionless subsystem.

The AI systems therefore served primarily as partners in exploring alternative viewpoints, verifying physical conjectures, and suggesting conceptual reformulations. However, the mathematical arguments, proofs, and physical interpretations presented here were developed and verified by the authors. Every theorem in this paper ultimately follows from explicit EFT assumptions, linear-algebraic arguments, anomaly cancellation, worldsheet current algebra, and the stated SCFT/LST classification inputs, independent of the conversational process through which the organizing ideas were discovered.

\subsection{Open problems}

\label{subsec:discussion-open-problems}

Although the present work establishes universal EFT bounds on all previously uncontrolled intersections between primitive BPS generators, several important questions remain open.

\begin{itemize}
    \item The EFT bounds obtained here are generally weaker than the corresponding bounds realized in F-theory. In particular, the isolated $(-3)$-generator satisfies the EFT bound $C_{-3}\cdot e\le3$, whereas the geometric bound is two. Similarly, some $(-2)$-generators in non-Higgsable clusters admit the EFT bounds $3$, $4$, $5$, and $7$, while F-theory realizes at most two~\cite{Morrison:2012np}. Understanding which additional quantum gravity consistency conditions are responsible for this gap remains an important open problem.

    \item The bound $C_{-2}\cdot e\le1240$ for an isolated gauged $(-2)$-generator is derived solely from the classification of affine embeddings into the $(E_8)_1$ current algebra~\cite{Dynkin:1957,Panyushev:2009,Panyushev:2014,DiFrancesco:1997nk}. It is almost certainly far from optimal. Determining the physically realizable maximum would require identifying which affine embeddings are compatible with the full 6d consistency conditions, including gauge anomaly cancellation, matter representations, global symmetry structure, and the complete tensor spectrum.

    \item From a mathematical viewpoint, it would be desirable to obtain closed expressions for the Zariski coefficients associated with arbitrary negative-definite trees and graphs. Such formulae would considerably simplify the computation of intersection bounds and make the lattice Zariski decomposition directly applicable to large-scale scans of allowed tensor charge lattices.

\item

More broadly, the lattice Zariski decomposition suggests that much of the tensor geometry may be reconstructed directly from low-energy physics. It would be interesting to combine anomaly cancellation, the attractor mechanism, lattice Zariski decomposition, and worldsheet current algebra into a unified algorithm for reconstructing the full primitive BPS cone, and ultimately the complete tensor sector and their classification, from EFT data alone~\cite{Kim:2024eoa,Hamada:2026zta}.

\end{itemize}

\bigskip

\acknowledgments
We would like to thank Yuta Hamada, Seongmin Jeon and Cumrun Vafa for helpful discussions. H.K. is supported by the National Research Foundation of Korea (NRF) grant funded by the Korean government (MSIT)
(2023R1A2C1006542).

\bibliographystyle{JHEP}
\bibliography{references}

\end{document}